# A Missing Key to Understand the Electrical Resonance and the Mechanical Property of Neurons: a Channel-Membrane Interaction Mechanism

*Shoujun Yu[1#], Tianruo Guo[2#], Wenji Yue[1], David Tsai[2], Yanlong Tai[1] Bing Song[2] and Hao Wang[1]*

[1]Institute of Biomedical & Health Engineering, Shenzhen Institutes of Advanced Technology (SIAT), Chinese Academy of Sciences (CAS), China

[2]Graduate School of Biomedical Engineering, University of New South Wales, Sydney, NSW2052, Australia

**Corresponding authors:**

Hao Wang (hao.wang@siat.ac.cn),

#These authors contributed equally to this manuscript.

**Abstract**

The recent study of the interaction between the fatty acyl tails of lipids and the $K^+$ channel establishes the connection between flexoelectricity and the ion channel's dynamics, named Channel-Membrane Interaction (CMI), that may solve the electrical resonance in neurons.

**A brief history of electrical resonance of neurons**

The cell membrane exhibits an inherent characteristic of frequency preference, similar to that of a resonance circuit [1]. Various electrical resonances have been observed in different phenomena, such as signal amplification at specific frequencies, frequency response resembling an RLC (resistor-inductor-capacitor) circuit, and voltage oscillation triggered by subthreshold stimulations. However, the complete understanding of the electrical resonance is still lacking, as there is currently no proper explanation for its physical origin. This commentary presents a potential biological explanation for the electrical resonance in neurons.

Figure 1(A1) illustrates a typical electrical resonance phenomenon: subthreshold oscillation observed when a subthreshold stimulus is applied [1]. This phenomenon can be accurately described by an RLC circuit [2], as shown in the figure.

**Two assumptions for the origin of resonance in neurons**

  a. **Flexoelectricity assumption—A mechanical perspective**

In the 1940s [3], Cole first observed the electrical resonance and proposed that it could be attributed to the piezoelectric effect of the cell membrane. With the discovery of the lipid bilayer structure of the cell membrane in the late 1950s, it became evident that this centrosymmetric double-layer dipole structure could generate flexoelectricity, which can be considered as a specialized form of piezoelectricity at the molecular level thickness [4] (see Figure 1(A2)). The cell membrane's intracellular and extracellular terminals carry negative charges, while the central parts carry positive charges. Thus, bending the cell membrane alters the distribution of negative charge on both sides, resulting in an electric field across the membrane, known as direct flexoelectricity. Additionally, an externally applied electric field can induce the bending of the cell membrane, known as reverse flexoelectricity. Flexoelectricity facilitates the conversion of energy between surface tension and the transmembrane electric field. Figure 1(B) provides a simple and qualitative model of flexoelectricity. The discrete distribution of positive and negative charges in the lipid bilayer can be observed in Figure 1(B1). By summing up the potentials generated by all charges along the x-axis crossing the membrane (Figure 1(B2)), the potential can be calculated. The two negative peaks and one positive peak correspond to the polarity distribution. When the membrane is bent, the concentration of negative charges on the intra- and extracellular surfaces changes in opposite directions, resulting in a difference in amplitude between the two negative peaks, known as $V_{outer-inner}$, a transmembrane potential. This potential's amplitude decreases as the bending radius decreases, as depicted in Figure 1(B3). **Further details of the modeling process can be found in Supplementary S1.**

This flexoelectricity generates the oscillatory response is explained as follow:

When an external electric field is applied to the cell membrane, it undergoes bending caused by reverse flexoelectricity. Due to the elastic properties of the cell membrane, it starts to vibrate [5-6]. During this vibration, the bending is converted into an oscillatory voltage through flexoelectricity, which can be measured as subthreshold oscillation.

However, a significant drawback of this assumption is its inability to explain why the resonance is significantly reduced when ion channels are blocked using pharmacological blockers [7].

  b. **Ionic assumption—An electrical perspective**

Another assumption suggests that the resonance arises from variations in the impedance of ion channels (Figure 1(A3)). In 1952, Hodgkin and Huxley proposed the Hodgkin-Huxley (H-H) model to replicate the action potential observed in giant squid axons [8]. The model incorporates an RC circuit and a set of equations that describe the dynamics of ion channels. It was proposed that the resonance could be attributed to changes in the conductance of potassium ($K^+$) channels. Subsequent studies revealed that not only $K^+$ channels, but also voltage-gated channels such as sodium ($Na^+$) and calcium ($Ca^{2+}$) channels, contribute to the electrical resonance [9]. Moreover, numerous experimental studies have reported the elimination of resonance when ion channels are blocked [10]. However, the precise mechanism underlying the resonance induced by ion channels remains unclear.

The main flaw in this assumption arises from the lack of a mechanism explaining the energy release by ion channels, which can manifest as negative impedance in a circuit. The ionic assumption is primarily supported by the H-H model, which mathematically generates subthreshold oscillation through the variation in ion channel impedance [8]. However, this impedance variation involves a negative impedance (the impedance decoupling process is illustrated in **Supplementary S2**). While the mathematical result is plausible, there is a lack of physical explanation for the negative impedance.

The issue lies in the fact that if a unit's impedance variation can generate oscillation, it must be capable of both consuming and producing energy. In circuit terms, this unit would generate both positive (energy-consuming) and negative (energy-producing) impedance. Similar to an RLC circuit, where the capacitor, C, can consume and produce energy through its electrical field, and the inductor, L, can consume and produce energy through its magnetic field, the capacitor and inductor exhibit positive and negative impedance, respectively. This principle can also be applied to the H-H model. Ion channels possess both positive and "negative impedance" to generate oscillation.

The crucial question is: How can ion channels, which consume energy, also produce energy (negative impedance)? This question holds the key to understanding the electrical resonance generated by ion channels.

**A Channel-Membrane Interaction (CMI): The mechanism of consuming and producing energy by ion channels**

Recent studies on the interaction between the fatty acyl tails of lipids and the K+ channel [11] have uncovered a connection between flexoelectricity and the dynamics of ion channels (Figure 1(A4)). This interaction, named as Channel-Membrane Interaction (CMI), may hold the key to unraveling the mystery of electrical resonance in neurons.

When the $K^+$ channel is opened by an applied voltage, the fatty acyl tails of lipids infiltrate and bond with the $K^+$ channel, resulting in surface tension as the $K^+$ channel pulls the inner layer of the cell membrane, causing it to bend upwards. The $K^+$ channel is capable of storing energy through the surface tension of the membrane. Once the gating process is completed, the $K^+$ channel releases the fatty acyl tails of lipids, allowing the membrane to recover its original bending shape and releasing the stored surface tension as energy. Additionally, due to flexoelectricity, the storage and release of surface tension can also be detected as voltage oscillation. The CMI mechanism enables the $K^+$ channel to store and produce energy. Furthermore, since other channels such as $Na^+$ and $Ca2^+$ also contribute to resonance activities, it is possible that the CMI mechanism is a prevalent mechanism in various types of ion channels.

Figure 2(A) provides a comprehensive illustration of the role of Channel-Membrane Interaction (CMI) in action potentials. In a typical action potential (Figure 2(A1)), the opening of $K^+$ channels (red curve) begins when the extracellular membrane potential is negative. At this stage, the cell membrane bends upward (Figure 2(A2-A3)) due to flexoelectricity. Simultaneously, the $K^+$ channel, through CMI, exerts a dragging force on the inner layer of the cell membrane during its opening, causing it to also bend upward (Figure 2(A4)). As a result, the CMI effect amplifies the impact of flexoelectricity, making the mechanical bending during the action potential more detectable. This cell membrane bending corresponds to an expanding mechanical pulse accompanying the action potential (Figure 2(A5)), as observed in Thomas Heimburg's study on lobster neurons [12].

In summary, the electrical resonance in neurons may arise from the flexoelectricity of the cell membrane. The CMI mechanism enhances the effect of flexoelectricity, thereby significantly amplifying the electrical resonance. Conversely, blocking ion channels can impede the CMI mechanism, leading to a substantial reduction in electrical resonance.

**A comparison of two assumptions**

The primary contribution of the CMI assumption is its ability to explain the resonance and oscillation observed in neurons. In conventional theory, the ionic current from ion channels is believed to be the exclusive source generating this resonance/oscillation, which is referred to as the ionic assumption. A comparison between these two assumptions can be made through a thought experiment involving the application of a DC subthreshold stimulus current, as depicted in Figure 2(B1).

According to the ionic assumption, the generated subthreshold oscillation would not decay as long as the stimulus to open ion channels is maintained. Therefore, the measured voltage/current oscillation would follow an undamped pattern, as shown in Figure 2(B2).

In contrast, the CMI assumption considers the oscillation to originate from the vibration of the cell membrane, which behaves as a lossy system with energy dissipation that can be modeled by an RLC circuit [2]. In this case, the oscillation can only persist for a limited duration, as illustrated in Figure 2(B4).

To investigate the damping characteristics of these oscillations as a potential distinguishing feature for their physical origins, we hypothesized and reviewed relevant references. It was found that both patterns of subthreshold oscillation can be observed in electrophysiology. For instance, pacemaker cells in cardiac myocytes exhibit sustained undamped oscillations dominated by the "funny" (If) current [13], as demonstrated in Figure 2(B3). This undamped subthreshold oscillation seems to align with the ionic assumption. However, in the peripheral nervous system (PNS) and central nervous system (CNS), a more frequently observed subthreshold oscillation follows a damping pattern (Figure 2(B4)). Therefore, the CMI assumption provides a more reasonable explanation for these damping patterns.

**Outlook-The linkage between mechanical and electrical properties of neurons**

It is known that the action potential exhibits a component of a mechanical wave. The soliton theory, pioneered by Thomas Heimburg [14], has shed light on this phenomenon. Similar to the wave-particle duality of light, the action potential may also exhibit an electrical-mechanical duality. The issue of action potential collision serves as an excellent example to illustrate this duality. According to saltatory conduction (an electrical characteristic), the collision of two action potentials would result in their annihilation. However, based on the mechanical characteristic, there is a small chance that these two action potentials could pass through each other [15]. Therefore, both the H-H model and the soliton theory may hold partial truths, and a multi-physics model that combines these aspects is yet to be established.

**Conclusion**

The Channel-Membrane Interaction (CMI) assumption, which involves the interaction between ion channels and the cell membrane via flexoelectricity, has provided a potential mechanism for ion channels to store and release energy during gating. This CMI mechanism offers a potential solution to the enigma of ion channel-induced electrical resonance observed in excitable cells. Additionally, it sheds light on the physical foundation of the mechanical wave observed in action potentials, which may result in crucial corrections to the conventional neural model. By uncovering the role of flexoelectricity and the CMI mechanism, a deeper understanding of the complex dynamics of excitable cells and the integration of electrical and mechanical aspects in neural signaling can be achieved.

**Acknowledgment**

This work was supported by grants from the National Natural Science Foundation of China (62071459), Guangdong Research Program (2019A1515110843), Shenzhen Research Program (GJHZ20200731095206018), and Shenzhen Science and Technology Program (JCYJ20220818101404009)

**Reference**

1. Mauro, A., Conti, F., Dodge, F. and Schor, R., 1970. Subthreshold behavior and phenomenological impedance of the squid giant axon. The Journal of general physiology, 55(4), pp.497-523.

2. Wang, H., Wang, J., Cai, G., Liu, Y., Qu, Y. and Wu, T., 2021. A physical perspective to the inductive function of myelin—A missing piece of neuroscience. Frontiers in neural circuits, 14, p.562005.

3. Cole, K.S., 1941. Rectification and inductance in the squid giant axon. The Journal of general physiology, 25(1), pp.29-51.

4. Petrov, A.G., 2002. Flexoelectricity of model and living membranes. Biochimica et Biophysica Acta (BBA)-Biomembranes, 1561(1), pp.1-25.

5. Petrov, A.G., Miller, B.A., Hristova, K. and Usherwood, P.N., 1993. Flexoelectric effects in model and native membranes containing ion channels. European biophysics journal, 22, pp.289-300.

6. Todorov, A., Petrov, A., Brandt, M.O. and Fendler, J.H., 1991. Electrical and real-time stroboscopic interferometric measurements of bilayer lipid membrane flexoelectricity. Langmuir, 7(12), pp.3127-3137.

7. Puil, E., Gimbarzevsky, B. and Spigelman, I., 1988. Primary involvement of K+ conductance in membrane resonance of trigeminal root ganglion neurons. Journal of neurophysiology, 59(1), pp.77-89.

8. Hodgkin, A.L. and Huxley, A.F., 1952. A quantitative description of membrane current and its application to conduction and excitation in nerve. The Journal of physiology, 117(4), p.500.

9. Lampl, I. and Yarom, Y., 1997. Subthreshold oscillations and resonant behavior: two manifestations of the same mechanism. Neuroscience, 78(2), pp.325-341.

10. Hutcheon, B. and Yarom, Y., 2000. Resonance, oscillation and the intrinsic frequency preferences of neurons. Trends in neurosciences, 23(5), pp.216-222.

11. Jin, R., He, S., Black, K.A., Clarke, O.B., Wu, D., Bolla, J.R., Johnson, P., Periasamy, A., Wardak, A., Czabotar, P. and Colman, P.M., 2022. Ion currents through Kir potassium channels are gated by anionic lipids. Nature communications, 13(1), pp.1-11.

12. Gonzalez-Perez, A., Mosgaard, L.D., Budvytyte, R., Villagran-Vargas, E., Jackson, A.D. and Heimburg, T., 2016. Solitary electromechanical pulses in lobster neurons. Biophysical chemistry, 216, pp.51-59.

13. DiFrancesco, D., 2010. The role of the funny current in pacemaker activity. Circulation research, 106(3), pp.434-446.

14. Appali, R., Petersen, S. and Van Rienen, U., 2010. A comparison of Hodgkin-Huxley and soliton neural theories. Advances in Radio Science, 8(BK. 1), pp.75-79.

15. Gonzalez-Perez, A., Budvytyte, R., Mosgaard, L.D., Nissen, S. and Heimburg, T., 2014. Penetration of action potentials during collision in the median and lateral giant axons of invertebrates. Physical Review X, 4(3), p.031047.

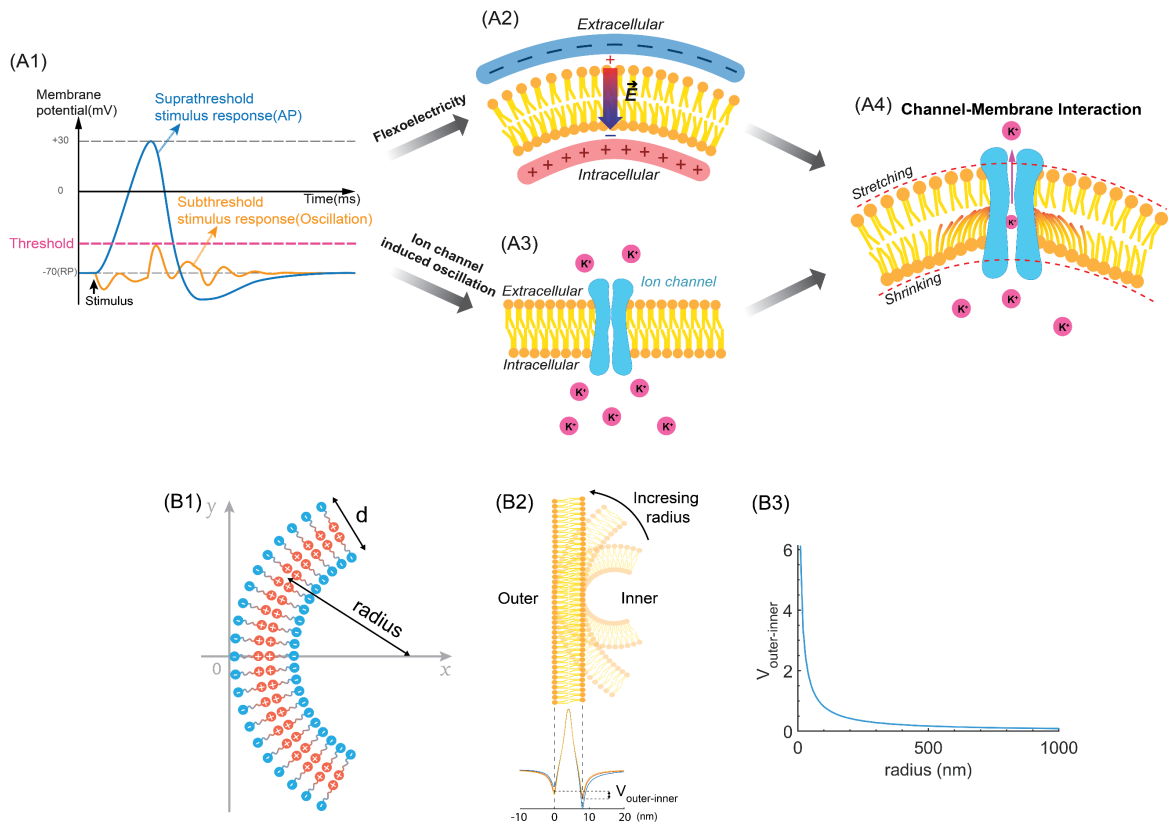

Figure 1. The phenomenon of electrical resonance in neurons. (A1) In response to external stimuli, a superthreshold stimulation triggers an action potential, whereas a subthreshold stimulation produces a voltage oscillation. Two assumptions have been proposed to explain this electrical resonance: the involvement of cell membrane flexoelectricity (A2) and the contribution of ion channels (A3). A recent discovery known as the channel-membrane interaction (CMI) suggests a more plausible explanation by combining these two assumptions (A4). (B) The concept of flexoelectricity in the cell membrane. The lipid bilayer structure of the membrane exhibits a curvature, with negatively charged intra- and extracellular terminals and positively charged intermediates (B1). As a result of the membrane curvature, the potential across the membrane displays two negative peaks at the terminals and a positive peak at the central position (B2). This curvature-induced potential difference, known as $V_{outer-inner}$, is a manifestation of flexoelectricity (B3), and its magnitude varies with the bending radius of the membrane.

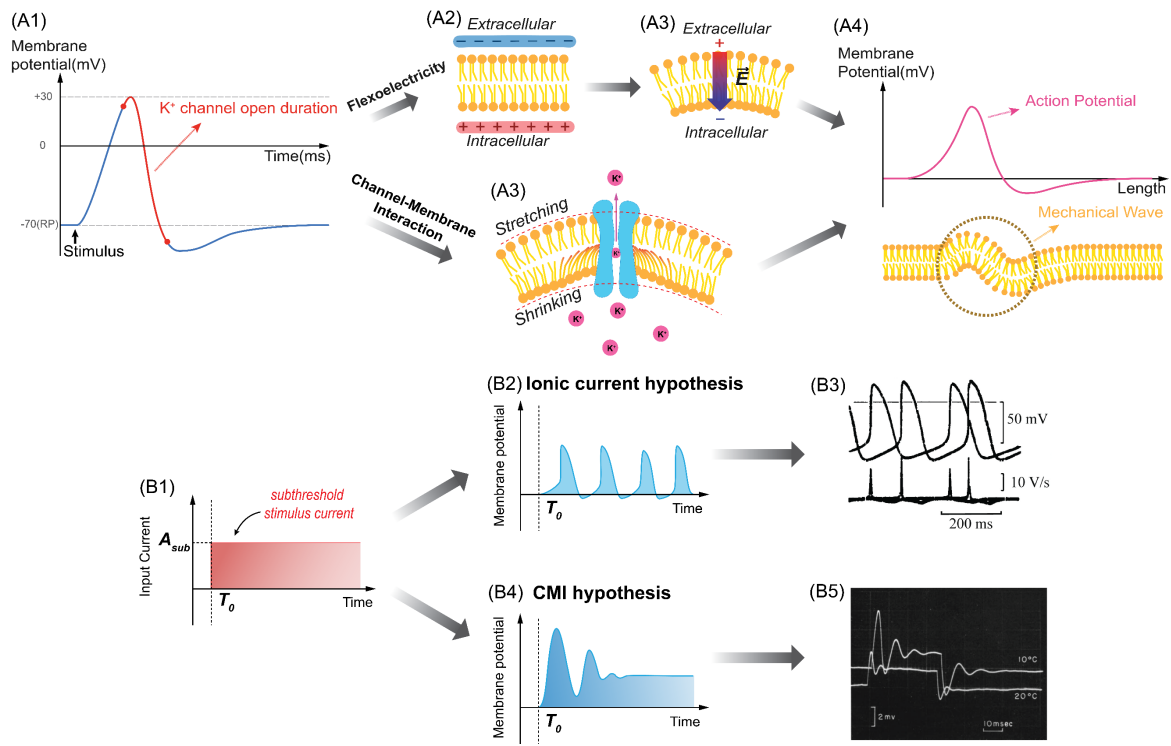

Figure 2. (A) The role of channel-membrane interaction (CMI) in action potentials. During the period of $K^+$ channel gating (A1), the membrane potential becomes extracellular negative and intracellular positive, causing the membrane to bend outward due to flexoelectricity (A2-A3). Additionally, CMI contributes to the outward bending of the cell membrane during $K^+$ channel gating (A4). As a result of the amplification provided by CMI, the membrane bending becomes a mechanical wave that accompanies the propagation of the action potential (A5). (B) The comparison between the two assumptions. In the sustained subthreshold stimulus current scenario (B1), the ionic assumption predicts a continuous oscillation pattern in the potential waveform (B2). Conversely, the CMI assumption suggests a damping oscillation pattern in the potential waveform (B4). The "funny" (If) current observed in pacemaker cells in cardiac myocytes exhibits a continuous oscillation pattern (B3), aligning with the ionic assumption. However, in other nervous systems, the measured subthreshold oscillation demonstrates a damping pattern (B5), supporting the CMI assumption.

# Supplementary

# A Missing Key to Understand the Electrical Resonance and the Mechanical Property of Neurons: a Channel-Membrane Interaction Mechanism

*Shoujun Yu[1#], Tianruo Guo[2#], Wenji Yue[1], David Tsai[2], Yanlong Tai[1] Bing Song[2] and Hao Wang[1]*

[1]Institute of Biomedical & Health Engineering, Shenzhen Institutes of Advanced Technology (SIAT), Chinese Academy of Sciences (CAS), China

[2]Graduate School of Biomedical Engineering, University of New South Wales, Sydney, NSW2052, Australia

**Corresponding authors:**

Hao Wang (hao.wang@siat.ac.cn),

#These authors contributed equally to this manuscript.

## S1. Modeling of flexoelectricity of cell membrane

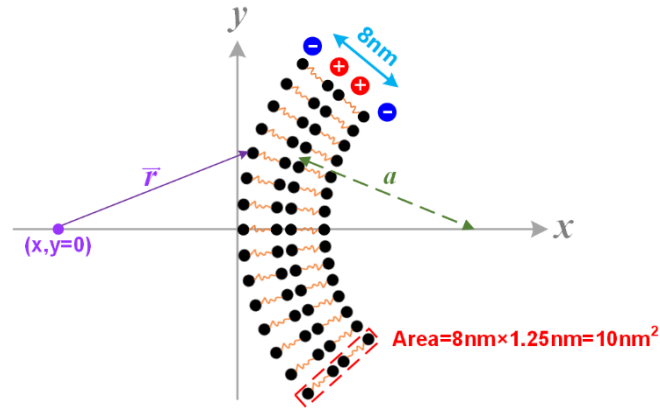

Figure S1. The modeling details of the calculation of the flexoelectricity of cell membrane

The detailed modeling process is illustrated in Figure S1. The arrangement of each polar is determined by the thickness of the bilayer and the diameter of the axon. Here axon radius, *a*, is a variable. The total thickness of the lipid bilayer is 8 nm, a typical value of cell membrane. The length of the dipole of each amphiphilic molecule is 3.6 nm, while the distance between the two positively charged polar is 0.8 nm. The cross-sectional area of each group of molecules is 10 nm². The charge quantity of the $n_{th}$ polar is $q_n$. The route from $n_{th}$ polar to the specific point (x,y=0) on the x-axis is $\vec{r_n}$. The total electric potential at the position (x,y=0) is the sum of the potential from each polar, as below:

$$\varphi = \frac{1}{4\pi\varepsilon_0}\sum\frac{q_n}{|\vec{r_n}|}$$

Here $q_n$ is a value with a sign corresponding to the polarity of the charge. Considering the value of each $q_n$ is identical, so

$$\varphi = \frac{1}{4\pi\varepsilon_0}\sum\frac{q_n}{|\vec{r_n}|} \propto \sum\frac{1}{|\vec{r_n}|}$$

So
$$\Delta V_{outer-inner} = \varphi|_{x=0\,nm} - \varphi|_{x=8\,nm}$$

## S2. Decoupling the negative impedance of ion channels in H-H model

1. **The comparison between the RLC circuit and the H-H model for decoupling the impedance of ion channels.**

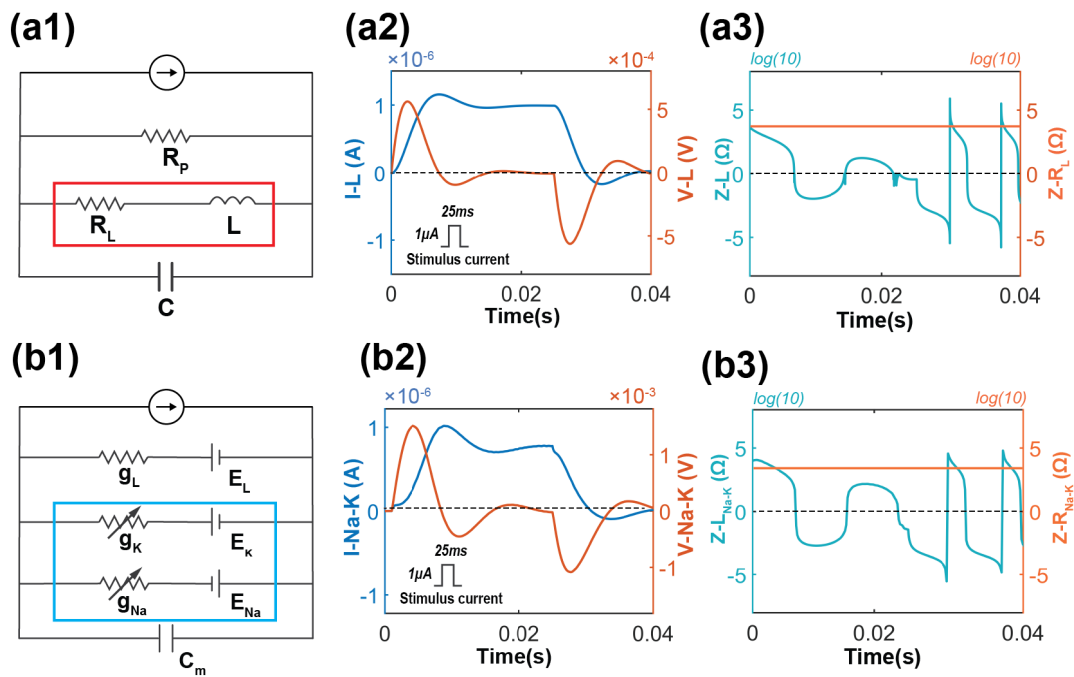

Figure S2.1. A comparison between the RLC circuit and H-H model under subthreshold stimulation conditions. (a1) The RLC circuit to duplicate the response of the H-H model in (b1); (a2) The voltage/current curves of the inductor branch of the RLC circuit (red box) by a square current stimulation; (a3) The impedance of the resistor, $R_L$ and the inductor, $L$, of the red box. (b1) The equivalent circuit of the H-H model; (b2) The voltage/current curves of the combined $K^+$ and $Na^+$ channel branches in the H-H model (blue box) by a square current stimulation; (b3) The impedance of the blue box can be decoupled as a resistor, $R_{Na-K}$, and an inductor, $L_{Na-K}$.

Figure S2.1(a) shows the modeling of a typical RLC circuit to illustrate the negative impedance issue. When a square wave current is applied to the RLC circuit, an oscillative voltage is generated due to the LC oscillation (Figure S2.1(a2)). Here the voltage /current of the inductor branch is depicted. The impedance ($V/I$) of the inductor branch, a resistor, $R_L$, and an inductor, $L$, are shown in Figure S2.1(a3). $R_L$ has a constant impedance, while $L$ shows changing impedance. As seen, an inductor can have a negative impedance. Physically, the positive impedance means the component absorbs energy, while the negative impedance means the component produces energy. The capacitor and inductor take turns storing (positive impedance) and releasing (negative impedance) energy, which is the origin of the LC oscillation.

A similar oscillation can also be observed in neurons when a subthreshold stimulation is applied. In this comment, we use the H-H model to demonstrate this oscillation, as shown in Figure S2.1(b). When a square wave subthreshold current is applied, the membrane voltage oscillates (Figure S2.1(b2)). Here we tune the parameter of the RLC circuit in Figure S2.1(a1) to produce the voltage waveform (Figure S2.1(a2)) the same as the one in the H-H model (Figure S2.1(b2)). The $K^+$ and $Na^+$ channel branches are considered as one branch (blue box) to show the current and voltage in Figure S2.1(b2). By comparing the RLC circuit and the H-H model, it is easy to find the sum of the $Na^+$ channel and the $K^+$ channel (blue box) shall represent an equivalent characteristic of the inductor branch (red box) in the RLC circuit, which are a constant impedance like $R_L$ and a changing impedance like $L$. We analyze the V-I curve to decouple the equivalent constant impedance, $R_{Na-K}$, and the changing impedance, $L_{Na-K}$, as shown in Figure S2.1(b3). **The detailed calculation process can be found in the following section.** In Figure S2.1(a3), $L_{Na-K}$ has a negative impedance following a real inductor's pattern, indicating that the ion channels can store and release energy.

## 2. Decouple the equivalent impedance of the Na-K channel in the H-H model

The following three major steps to decouple the equivalent impedance of the ion channel branches in the H-H model are listed.

a. Preprocess the modeling data of the H-H model.
b. Calculate the equivalent resistor $R_{Na-K}$.
c. Calculate the impedance of $L_{Na-K}$.

### a. Preprocess the modeling data of the H-H model.

The original result of the H-H model cannot be directly used for the impedance calculation due to the non-zero initial current, as shown in Figure S2.2. When a subthreshold square current pulse is applied, the current curves of $Na^+$ and $K^+$ channel branches are shown in Figure S2.2(a). These current curves do not increase from zero, meaning there is a DC bias in the current when no stimulus is applied. This DC biasing current is mainly to maintain the resting potential in the H-H model and shall not be attributed to the property of the voltage-gated $Na^+$ and $K^+$ channels in reality. Thus, we need to remove this DC biasing current to start these current curves from zero, as shown in Figure S2.2(b).

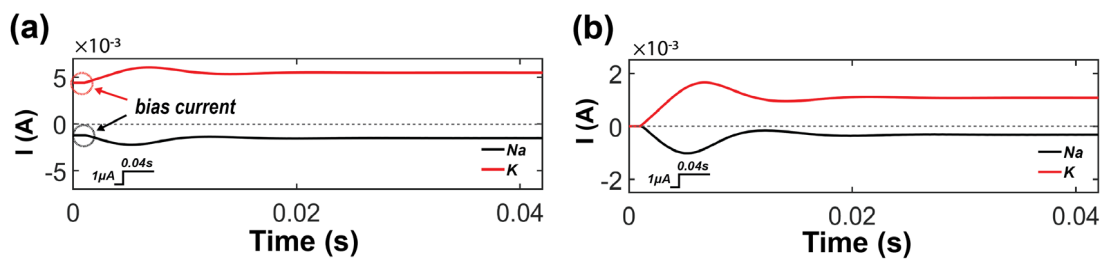

Figure S2.2 Illustration of preprocessing of the current in the $Na^+$ and $K^+$ branches. (a) The original current in the $Na^+$ and $K^+$ channel branches. (b) The rectified current without bias in the $Na^+$ and $K^+$ channel branches.

### b. Calculate the equivalent resistor $R_{Na-K}$

To calculate $R_{Na-K}$, we need to explain the concept of $V_{DC}$ and $I_{DC}$ by modeling of RLC circuit.

The modeling result of the RLC circuit by applying a continuous step current is shown in Figure S2.3 (a1). This continuous step current has an infinite duration, which can be considered a DC current except for the initial rising edge. Thus, the oscillation only happens at the initial stage. The voltage and current curves of the inductor branch will finally reach constant values, which are labeled as $V_{DC}$ and $I_{DC}$ in Figure S2.3 (a1). In this DC current duration, the RLC circuit can be simplified, as shown in Figure S2.3(a2). In DC bias, the inductor is a short circuit (SC), and the capacitor is an open circuit (OC). Therefore, the RLC circuit is reduced to just two resistors. The relationship between $V_{DC}$ and $I_{DC}$ is

$$R_L = \frac{V_{DC}}{I_{DC}}$$

Then the voltage on $R_L$ can be calculated as

$$V_{RL} = R_L \times I_{L-RL}$$

Since the total voltage of the inductor branch, $V_{L\text{-}RL}$, is the sum of the voltage on the inductor, $V_L$, and the voltage on the resistor, $V_{RL}$ (Figure S2.3(a4)), we can obtain the inductor's voltage by $V_L = V_{L-RL} - V_{RL}$, as shown in Figure S2.3 (a4). Then the impedance of the inductor can be calculated as $Z_L = \frac{V_L}{I_L}$, as shown in Figure S2.3 (a5).

The same process can be applied to the modeling results of the H-H model. Here we consider the $Na^+$ and $K^+$ channel branches as one branch by adding the two current curves in Figure S2.2(b) together as $I_{Na\text{-}K}$ in Figure S2.3(b1). It can be observed that there are also $V_{DC}$ and $I_{DC}$ in the voltage and current curves. The existence of $V_{DC}$ and $I_{DC}$ means there is also a resistor in this branch named $R_{Na\text{-}K}$, which calculated as

$$R_{Na-K} = \frac{V_{DC}}{I_{DC}}$$

The voltage of $R_{Na\text{-}K}$ can be calculated as

$$V_{R_{Na-K}} = R_{Na-K} \times I_{Na-k}$$

### c. Calculate the impedance of $L_{Na\text{-}K}$.

The existence of $R_{Na\text{-}K}$ means that this branch has at least two components. One is $R_{Na-K}$, another is some unknown component named $Z_{unknown}$, as shown in Figure S2.3(b2). We can calculate the voltage on $Z_{unknown}$ by $V_{unknown} = V_{Na-K} - V_{R_{Na-K}}$, as shown in Figure S2.3(b3). Then the impedance of $Z_{unknown}$ is calculated as

$$Z_{unknown} = \frac{V_{unknown}}{I_{Na-k}}$$

The result is shown in Figure S2.3(b4). The curve shares a similar shape as the one in Figure S2.3(a5), which is an inductor's impedance shape. Therefore, we can claim that the $Na^+$ and $K^+$ channel branches can be modeled as an inductor connected in series with a resistor.

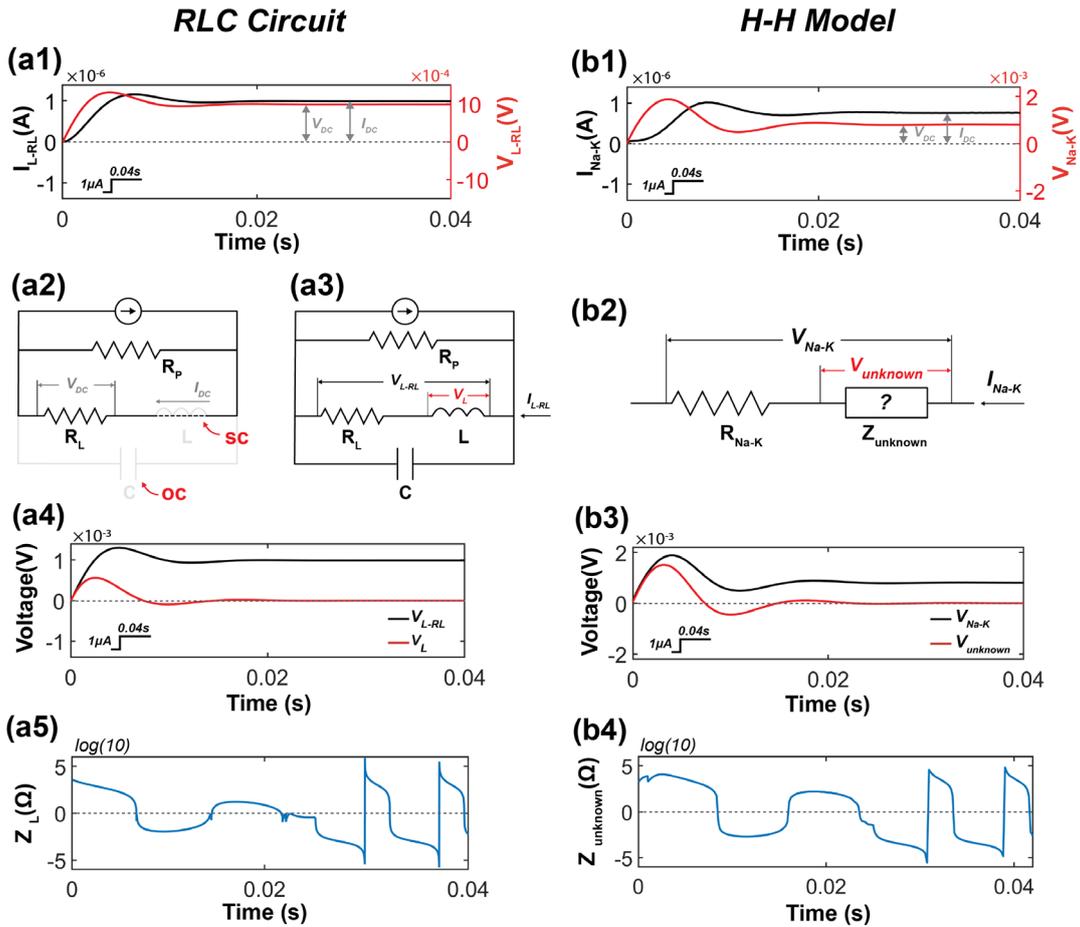

Figure S2.3 A comparing analysis of compositions between (a) the RLC circuit and (b) the H-H model. (a1) The voltage/current curves of the inductor branch of the RLC circuit. (a2) The simplified circuit of the RLC circuit at DC. (a3) The RLC circuit at general condition. (a4) The voltage of the inductor branch and the voltage of L. (a5) The changing impedance of L. (b1) The voltage/current curves of the Na$^+$ and K$^+$ channel branches in the H-H model. (b2) The equivalent branch of the Na$^+$ and K$^+$ channel branches in the H-H model (b3) The voltage of the equivalent branch and the voltage on the equivalent $Z_{unknown}$. (b4) The equivalent changing impedance $Z_{unknown}$.